\documentclass[aps,showpacs,twocolumn,pra,notitlepage,nofootinbib,superscriptaddress,10pt,floatfix]{revtex4-2}

\usepackage{graphicx} 
\usepackage{amsfonts}
\usepackage{amssymb}
\usepackage{amscd}
\usepackage{amsthm}
\usepackage{bbm}
\usepackage{booktabs} 
\usepackage{enumitem}

\usepackage[newcommands]{ragged2e}
\usepackage{floatrow}
\usepackage[caption=false, justification=centerlast, label font=bf]{subfig}
\usepackage{silence}
\usepackage{xcolor}
\usepackage[all]{xy}
\usepackage{physics}
\usepackage{appendix}
\usepackage{algpseudocode,algorithm}
\usepackage{mathtools}
\usepackage{comment}

\usepackage[colorlinks, linkcolor=red, anchorcolor=blue, citecolor=green]{hyperref}
\usepackage{fancyvrb}
\usepackage{tabularx}

\graphicspath{{./Figures/}}

\makeatletter
\long\def\@makecaption#1#2{%
  \vskip\abovecaptionskip
  \sbox\@tempboxa{#1: #2}%
  \ifdim \wd\@tempboxa >\hsize
    \justifying\noindent #1: #2\par
  \else
    \global \@minipagefalse
    \hb@xt@\hsize{\hfil\box\@tempboxa\hfil}%
  \fi
  \vskip\belowcaptionskip}
\makeatother

\begin{document}
\title{Entanglement detection in dynamical quantum systems}
\author{Xingjian Zhang}
\thanks{Present address: Centre for Quantum Software and Information, University of Technology Sydney, Sydney, Australia}
\affiliation{QICI Quantum Information and Computation Initiative, School of Computing and Data Science, The University of Hong Kong, Pokfulam Road, Hong Kong}
\author{Jiayi Wu}
\affiliation{QICI Quantum Information and Computation Initiative, School of Computing and Data Science, The University of Hong Kong, Pokfulam Road, Hong Kong}
\author{Jue Xu}
\affiliation{QICI Quantum Information and Computation Initiative, School of Computing and Data Science, The University of Hong Kong, Pokfulam Road, Hong Kong}
\author{Qi Zhao} 
\email{zhaoqi@cs.hku.hk}
\affiliation{QICI Quantum Information and Computation Initiative, School of Computing and Data Science, The University of Hong Kong, Pokfulam Road, Hong Kong}
\date{\today}

\begin{abstract}
    Quantum science and engineering increasingly target the exploitation and control of large-scale dynamical systems. As system size grows, novel phenomena and quantum advantages emerge, but the resulting many-body dynamics become correspondingly harder to monitor and characterize. In particular, the underlying Hamiltonian often lacks symmetries, and the induced many-body dynamics quickly become hard to track. Focusing on one of the most important quantum features---entanglement---we demonstrate that, for a broad class of common Hamiltonian models, one can design effective witnesses that detect both the presence and structure of entanglement over relatively long evolution times. Inspired by stabilizer-based simulation, we track the evolution of generalized stabilizers of the underlying state and use them to construct linear entanglement witnesses. Through numerical experiments on Heisenberg-type Hamiltonians with up to twelve qubits, we show that the required number of measurement settings can be kept at a modest scale as the system size increases, and entanglement detection remains effective over evolution times that significantly exceed those accessible to naive approaches. The corresponding calculations can be carried out largely symbolically. Furthermore, unlike accurate state simulation, the stabilizer-based design of entanglement witnesses can be performed in a much more coarse-grained, digitized manner. We hope these results will stimulate further exploration of the quantum control of large systems.
    
\end{abstract}

\maketitle

\emph{Introduction. ---}
In quantum science and engineering, increasing the system size often leads to qualitatively new phenomena and functionalities. Anderson's saying ``more is different'' encapsulates the idea that certain collective quantum behaviors emerge only in the many-body regime \cite{anderson1972more}. Prominent examples include topological phases and phase transitions  \cite{martinezRealtimeDynamicsLattice2016, landsmanVerifiedQuantumInformation2019,bernienProbingManybodyDynamics2017,ebadiQuantumPhasesMatter2021,wangRealizationFractionalQuantum2024, guoSiteresolvedTwodimensionalQuantum2024,abaninObservationConstructiveInterference2025}, whose characteristic long-range entanglement cannot be understood as a simple aggregation of the properties of smaller subsystems. In quantum information science, scaling up the system is likewise essential for a broad range of information-processing tasks. In addition to the sheer need for a large number of qubits to implement quantum error correction \cite{gottesman1997stabilizer,terhal2015quantum} and practically useful quantum algorithms \cite{montanaro2016quantum}, larger systems naturally give rise to rich structures of multipartite entanglement \cite{guhneEntanglementDetection2009}. 
Particularly, entanglement is crucial to the quantum advantage via quantum simulation \cite{ciracGoalsOpportunitiesQuantum2012,zhaoEntanglementAcceleratesQuantum2025,kingBeyondclassicalComputationQuantum2025} and indispensable for networked applications \cite{wehner2018quantum,brassard2007anonymous,defienne2024advances}.

In the quest to explore and exploit large quantum systems, a pivotal prerequisite is the development of effective methods to characterize and monitor their evolution. Although state and process tomography provide a complete description of a quantum system \cite{raymer1994complex,leonhardt1996discrete,leonhardt1997measuring,poyatos1997complete,chuang1997prescription,ariano2001quantum}, they require resources that scale exponentially with system size \cite{haah2016sample,donnell2016efficient1,donnell2016efficient2}, rendering them prohibitively expensive beyond a modest number of qubits. 
In many situations, however, it is not necessary to reconstruct the full state or process: learning only selected properties is sufficient to draw conclusions or to accomplish a given task. As mentioned earlier, one such central property

Experimentally, efficient entanglement detection has been demonstrated in systems comprising tens of qubits \cite{monz2011qubit,friis2018observation,cao2023generation}, a regime large enough to support certain practically relevant information-processing tasks.

However, in most reported experiments, the underlying entangled states---such as GHZ and W states---are highly symmetric and generated by artificially engineered processes over relatively short timescales. In contrast, in more general settings such as the study of generic many-body Hamiltonians, the dynamics typically exhibit only limited symmetries, and the resulting states become complex and difficult to track after relatively long evolution. Although such states are expected to be highly entangled almost surely \cite{hayden2006aspects,gross2009most}, it is not obvious whether their entanglement can be efficiently detected in practice. In particular, entanglement detection in large systems without proper prior information is almost impossible \cite{liu2022fundamental}.

In this work, we introduce a general scheme to efficiently detect entanglement generated by generic Hamiltonian dynamics within reasonable evolution times. Our method is inspired by stabilizer-based simulation \cite{gottesman1997stabilizer}: we track the evolution of a set of generalized stabilizers and use them to construct linear (genuinely multipartite) entanglement witnesses (EWs) tailored to the evolved state. A key observation is that, for the sole purpose of entanglement detection, the stabilizer dynamics need only be simulated approximately, in a coarse-grained, digitized manner. Furthermore, the approximate stabilizer evolution can be handled in an almost fully symbolic way, which substantially reduces the classical computational overhead. In addition to analytical analysis, we demonstrate the practicality of our approach on concrete Hamiltonian models via numerical simulation, which provides experimentally feasible EWs over relatively long times with a modest number of measurement settings.

\emph{Preliminaries. ---}
Before presenting our results, we review several necessary notions. The central notion is entanglement. Consider a bipartite quantum state $\rho\in\mathcal{D}(\mathcal{H}_A\otimes\mathcal{H}_B)$. In this work, we are always interested in entanglement over a given subsystem partition. Instead of performing full state tomography, one may apply a resource-efficient linear entanglement witness (EW) to detect entanglement in the system \cite{horodecki2001separability,bruss2002reflections}. Given a Hermitian operator $W$, it is an EW if $\tr(W\sigma) \geq 0$ for every separable state $\sigma$, while there exists at least one entangled state $\rho$ such that $\tr(W \rho)<0$. For a given target entangled state $\rho$, the optimal EW for its detection against noise is $W = \alpha I - \rho$, with $\alpha=\max_{\sigma\in\mathrm{SEP}}\tr(\sigma\rho)$ over the set of all separable states $\mathrm{SEP}$. However, it is challenging to describe the boundary of $\mathrm{SEP}$ in an explicit and accurate manner \cite{gharibian2010strong}. 
Instead, we may relax the problem by requiring $\tr(W\sigma)\geq 0$ for a convex superset of $\mathrm{SEP}$, $\mathcal{T}\supseteq\mathrm{SEP}$, where $\mathcal{T}$ is chosen to be sufficiently tight while admitting a simple characterization. One such relaxation is the set of states with a positive partial transpose (PPT). A bipartite state $\rho\in\mathcal{D}(\mathcal{H}_A\otimes\mathcal{H}_B)$ is a PPT state if its partial transpose with respect to $\mathcal{H}_B$, $\rho^{\mathrm{T}_B}$, remains positive semi-definite, $\rho^{\mathrm{T}_B} \geq 0$. 

When a quantum system comprises more than two subsystems, we can further discuss states with genuine multipartite entanglement (GME). Generally, consider a tripartite quantum state  $\rho$ acting on $\mathcal{H}_A\otimes\mathcal{H}_B\otimes\mathcal{H}_C$. We call it a bi-separable state if it admits a convex decomposition of states, $\rho=\sum_ip_i\rho_i$, where $p_i$ gives a probability distribution, and $\rho_i$ is separable with respect to one of the partitions $A|BC,B|AC$, or $C|AB$. Otherwise, we call it a GME state. Similarly, we can construct linear witnesses to detect the existence of GME.

To construct linear EWs for the quantum states of interest, we shall utilize the generalized stabilizer formalism \cite{gottesman1997stabilizer,gottesmanHeisenbergRepresentationQuantum1998}. Consider a general $n$-qubit operator with eigenvalues $\pm1$, $S$. We say a state $\psi$ is stabilized by $S$ if $S\ket{\psi} = \ket{\psi}$, and we call $S$ a stabilizer of $\ket{\psi}$. It can be proven that given $n$ mutually commuting and independent $n$-qubit operators of this type, $\{S_1,\cdots,S_n\}$, they uniquely determine an $n$-qubit pure quantum state $\ket{\psi}$. Furthermore, for every $n$-qubit pure quantum state, we can find $n$ such operators that stabilize it. 
Since the set of Pauli operators $\mathbb{P}$ defines a complete basis of the operator space, we can describe the generalized stabilizer operators as a linear combination of Pauli strings:
\begin{equation}\label{eq:genstab}
    S_i=\sum_j\lambda_j\bigotimes_{k=1}^{n}\sigma_j^{(k)},\sigma_j^{(k)}\in\mathbb{P}.
\end{equation}
As a direct corollary, the projector onto $\ket{\psi}$ can be written as $\ketbra{\psi}=\prod_{i = 1}^n (1 + S_i)/2$.

\emph{Algorithm to design the EW---}
We now describe the problem setting. Consider a realistic quantum platform that implements a prescribed Hamiltonian evolution on $n$ qubits, specified by an initial quantum state $\ket{\psi(0)}$, a Hamiltonian operator $H$, and an evolution time $t$. The system might suffer from noise and loss, but the actual system evolution should not significantly deviate from $\ket{\psi(t)}=U\ket{\psi(0)}$ with $U=\exp(-iHt)$. Our goal is to detect whether the evolved state is (genuinely multipartite) entangled, across a given partition of qubits using a reasonable number of measurement settings, ideally with polynomial or sub-exponential scaling in $n$, i.e., $2^{o(n)}$. 

For this purpose, we consider using the generalized stabilizers of the evolved state to construct (multipartite) EWs. As mentioned earlier, while the operator of $W=\alpha I-\ketbra{\psi(t)}$ is a (nearly) optimal linear EW with respect to $\ket{\psi(t)}$ that is robust to noise, it is generally not easy to measure. In a realistic experiment, one needs to decompose $W$ into simpler observables that can be measured easily. Here, we consider implementing the EW by measuring Pauli strings and carrying out relevant classical data post-processing, which represents a natural choice for many experimental platforms. As such, we describe the number of measurement settings or the EW size as the number of Pauli strings to be measured.

To obtain an EW requiring fewer measurement settings, we consider relaxing the EW design to a less robust but simpler one. Heuristically, the operator $W'=\beta I-\sum_{i=1}^{n}S_i$ often serves as a good EW with a suitable $\beta$, particularly when the state is a stabilizer state \cite{tóth2005entanglement}. This EW is usually easier to measure than $W$ as it is usually decomposed into fewer terms in the Pauli basis. Note that some Pauli strings, such as $Z\otimes I$ and $I\otimes Z$, can be measured simultaneously bacause they commute, hence the number of measurement settings can be further reduced compared to the number of Pauli strings in the decomposition. 

To this end, we are left with the question of determining $\sum_{i=1}^{n}S_i$ associated with the target state. Since we are aware of the initial state before evolution, $\ket{\psi(0)}$, we can determine its generalized stabilizers, $S_i^{T=0}$. We shall use the superscript to denote the evolution time and simply abbreviate the stabilizer as $S_i^{(t)}$ for $T=t$. At time $T=t$, the generalized stabilizers evolve into $U S_i^{(0)}U^\dag$. However, $U$ might be hard to compute for long evolution times. For this purpose, we further approximate the evolution via digitization. By dividing the total time $t$ into $r$ steps and approximating the short-time evolution of each step using the second-order Taylor expansion, we approximate the Hamiltonian evolution in time $t$ as
\begin{equation}
    U\approx\tilde{U}=\left[I - iH\frac{t}{r} - \frac{1}{2} H^2 \left(\frac{t}{r}\right)^2\right]^r.
\end{equation}
Combining the above steps, we are able to perform symbolic calculation and approximate each stabilizer as $S_i^{(t)}\approx\tilde{S_i}^{(t)}\equiv\tilde{U}S_i^{(0)}\tilde{U}^\dag$. Because of linearity, we may simply update $\sum_{i=1}^{n}\tilde{S_i}^{(t)}$ rather than updating each stabilizer independently. Finally, by solving an SDP to determine the EW parameter $\beta$ with respect to $\mathrm{PPT}$ states, we obtain a valid linear EW.

So far, the calculation is similar to the stabilizer-based Hamiltonian simulation. Nevertheless, we observe the following facts in later numerical experiments:
\begin{enumerate}[leftmargin=*,parsep=0pt]
    \item 
    Regarding the approximation error, we typically have
    \begin{equation}
        \left|\sum_{i=1}^{n}S_i^{(t)}-\sum_{i=1}^{n}\tilde{S_i}^{(t)}\right|<\sum_{i=1}^{n}\left|S_i^{(t)}-\tilde{S_i}^{(t)}\right|,
    \end{equation}
    and the difference between the two sides of the inequality can be large when the system has high symmetry.
    \item
    In the decomposition of $\sum_{i=1}^{n}\tilde{S_i}^{(t)}$, some Pauli strings are associated with coefficients whose absolute values are extremely small.
\end{enumerate}
In the first observation, generally, various coefficients in the Pauli-basis decomposition of the generalized stabilizers may take opposite signs; hence the approximation errors in each evolved stabilizer cancel out to some extent when summed up. As a consequence for designing an EW, intuitively, we can take a much coarser approximation in updating the stabilizers by dividing the evolution time into fewer steps than for usual Hamiltonian simulation. The second observation suggests that we may neglect the less significant Pauli-basis terms in the decomposition of stabilizers. Following the two observations, in addition to using a coarser time-step division, we further take the following steps in designing the EW:
\begin{itemize}[leftmargin=*,parsep=0pt]
    \item Step 1: After a unit time $t_0$, decompose $\sum_{i=1}^{n}\tilde{S_i}^{(t)}$ in the Pauli basis and discard the terms whose coefficients are below a given threshold $l$;
    \item Step 2: Update $\sum_{i=1}^{n}\tilde{S_i}^{(t_0)}$ for another unit time $t_0$ via digitized Hamiltonian evolution.
\end{itemize}
We can repeat the above two steps until a pre-determined target time $t_F$ for detecting entanglement. We summarize our algorithm for designing an EW in Appendix A.

\emph{Numerical results. ---}
Using the algorithm described above, we design EWs to detect (genuine multipartite) entanglement at a general time during Hamiltonian evolution. We consider several representative Hamiltonian models in the Heisenberg family.

\begin{figure}[hbt!]
\centering
\includegraphics[width=0.83\textwidth]{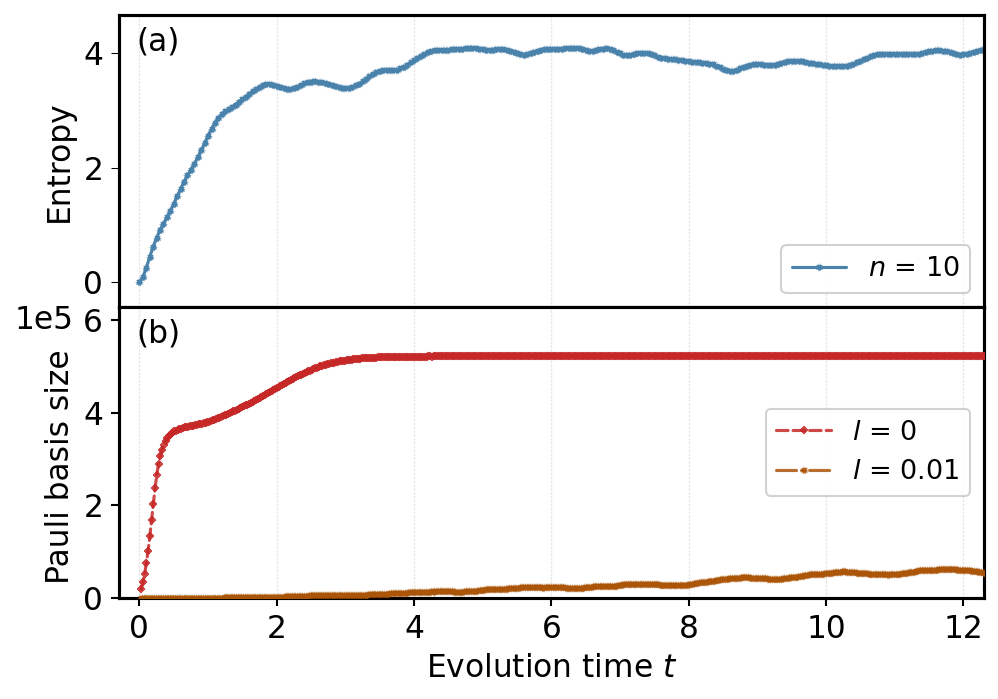} 
\caption{Evolution features of $H_{XY,\mathrm{power}}$ in Eq.~\eqref{eq:powerlaw} with a system size of 10 qubits. 
(a) Evolution of the entanglement entropy. 
(b) Size of the optimized EW for the evolving state. We digitize $t_0$ into $r$ steps and carry out simulation with ($l=0.01$) and without ($l=0$) truncation of Pauli basis terms with small coefficients.}
\label{Fig:EntEvo}
\end{figure}

\emph{Showcase: $XY$-Heisenberg model.---}
The first Hamiltonian model we consider is the $XY$-Heisenberg model with transverse interaction strength obeying a power-law decay and a longitudinal field:
\begin{equation}\label{eq:powerlaw}
    H_{XY,\mathrm{power}}=\sum_{i}^{n}J_{i j}(X_iX_j+Y_iY_j)+B_Z\sum_i^nZ_i,
\end{equation}
where we denote the three non-identity Pauli operators on the $i$'th qubit as $X_i,Y_i$ and $Z_i$, the correlation between the $i$'th and the $j$'th qubits is subjected to a power-law decay distribution, $J_{ij}=|i-j|^{-2}$, and a longitudinal field is present with $B_Z=0.2$. In addition, we consider a periodic boundary condition. The initial state in the evolution is $\ket{\psi_0}=\ket{01010\cdots}$, where the qubits with odd indices are initialized in $\ket{0}$ and stabilized by $Z$, and those with even indices are initialized in $\ket{1}$ and stabilized by $-Z$. For simplicity, we investigate entanglement between the set of qubits labeled from $1$ to $\lfloor n/2\rfloor$ and the set of remaining qubits. We showcase the entanglement evolution of this Hamiltonian model in Fig.~\ref{Fig:EntEvo}(a) for a ten-qubit instance. We quantify system entanglement by entanglement entropy, defined as $-\tr(\rho_A\log\rho_A)$ for a system partition of $A$ and $\bar{A}$. It can be seen that entanglement soon increases as the state evolves and remains at a relatively high value for a long time.

\begin{figure*}[t]
    \centering
    \includegraphics[width=0.83\textwidth]{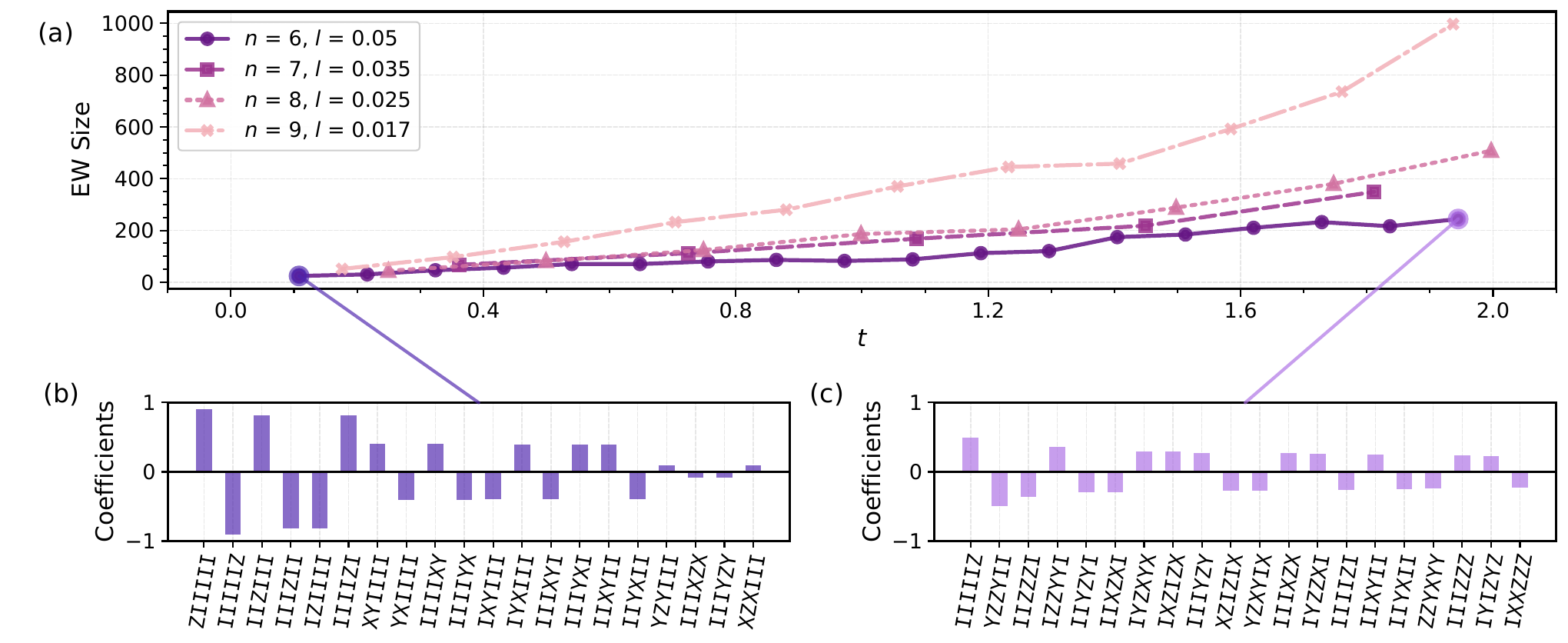}
    \caption{Evolution of the EW features under the $H_{XY,\mathrm{power}}$ Hamiltonian model. (a) EW-size evolution under different system sizes. We perform simulation with a system size ranging from $6$ to $9$ qubits, and the truncation threshold values are heuristically optimized accordingly. Simulation stops at the time when we cannot detect entanglement using the optimized EW.
    (b) The 20 coefficients with the largest absolute values in the Pauli-basis decomposition of the EW at time $t=t_0$ (showcased for six qubits), along with their corresponding Pauli strings.
    (c) The 20 largest Pauli decomposition coefficients of the EW at a later evolution time $t=18t_0$.}
\label{Fig:EWSizeEvo}
\end{figure*}

To reduce the experimental resources for entanglement detection, we employ a relatively coarse digitization when simulating the stabilizer evolution. In Supplementary Material (SM)~Sec.~II, we analytically derive the digitization error of $S_i$ in a short time and provide numerical evidence validating its effectiveness. Parallelly, we implement the truncation technique that repeatedly discards the least significant terms in the Pauli-basis decomposition of the approximate stabilizers. As indicated by the brown dotted line in Fig.~\ref{Fig:EntEvo}(b), by setting a truncation threshold of $l=0.05$, the EW sizes after truncation are significantly smaller than those obtained when all terms are kept. We further investigate the efficacy of this truncation in Fig.~\ref{Fig:EWSizeEvo}. For each system size, the truncation thresholds are heuristically adjusted to maximize the time window during which an effective EW can be successfully constructed for the evolving state. Within the algorithm, we define a unit time $t_0 = 1/{2^{\log^2 n}}$ discretized into $r=10$ steps. Fig.~\ref{Fig:EWSizeEvo}(a) plots the evolution of the output EW sizes throughout the Hamiltonian dynamics. Each curve terminates when the algorithm can no longer generate an EW that successfully detects entanglement in the ideal system. These trajectories demonstrate that the truncation technique effectively bounds the EW size at a moderate, experimentally manageable level.

As certain information about the evolving system's stabilizers is lost in the digitization and truncation, the algorithm generally fails at long times. To investigate this, we examine the Pauli-basis decomposition coefficients of the EW. Fig.~\ref{Fig:EWSizeEvo}(b) and (c) showcase the 20 most significant coefficients for a six-qubit system at $t_0$ and at the simulation stopping point, respectively. At $t_0$, the coefficient distribution is highly peaked. Beyond the 20 depicted terms, the remaining of the $808$ total coefficients have absolute values below $0.017$, indicating that a few dominant Pauli strings characterize the evolved stabilizers during short-time evolution. As the system evolves, this distribution gradually flattens [Fig.~\ref{Fig:EWSizeEvo}(c)]. Consequently, the fixed truncation threshold becomes overly aggressive, discarding critical information required for entanglement detection and ultimately limiting long-time verification. We discuss the quantitative impact of this truncation threshold in SM~Sec.~III.

\begin{figure*}[t]
\centering
\includegraphics[width=0.83\textwidth]{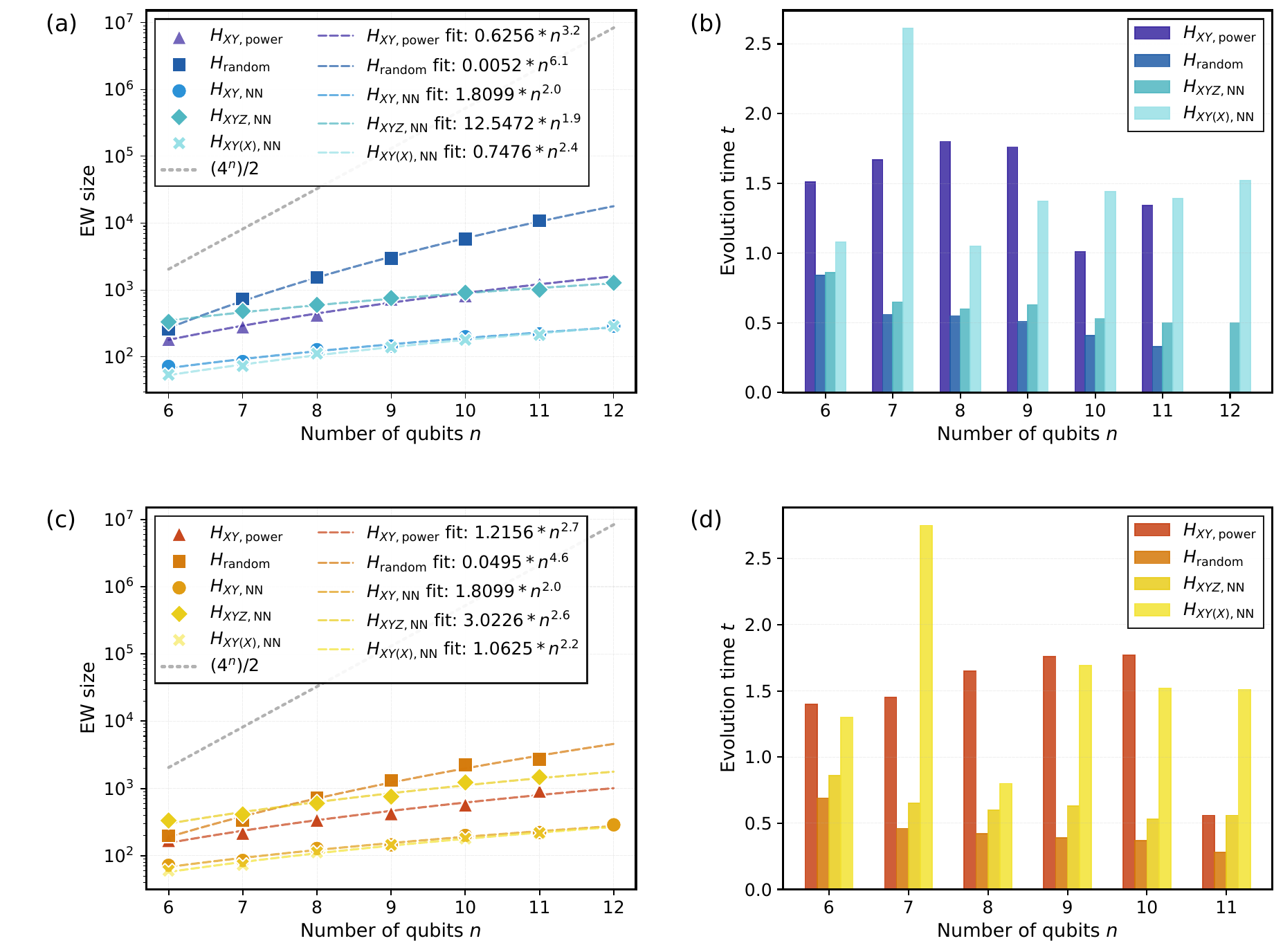}
\caption{EW performance with respect to the system size for five models. Under the constraint that the EW size scales at most polynomially with the system size, we depict the EW size (curves) and the maximal entanglement detection time (bars) in the optimization for bipartite entanglement in (a-b) and GME in (c-d), respectively. The maximal entanglement detection times for $H_{XY,\mathrm{NN}}$ are provided separately in Fig.~\ref{fig:XYNNEWsize}.}
\label{fig_poly_rela_2p}
\end{figure*}

Finally, Fig.~\ref{fig_poly_rela_2p} demonstrates the performance of the heuristically optimized EW. To ensure practical experimental implementation, the EW size is constrained to scale polynomially with the system size $n$. Under this constraint, we evaluate both the duration of valid entanglement detection and the EW size at the stopping point. As shown by the blue triangles in Fig.~\ref{fig_poly_rela_2p}(a), the EW size grows much more slowly than the total number of terms in the Pauli basis, requiring significantly fewer measurement settings than full state tomography. Furthermore, as shown by the blue bars in Fig.~\ref{fig_poly_rela_2p}(b), the EW successfully detects entanglement for relatively long times, with the maximum detection time showing no significant degradation as the system size increases.

\emph{General Hamiltonian models. ---}
To further demonstrate the applicability of our method, we follow a similar approach to designing an EW for several other Hamiltonian models. These include (1) the transverse-field Ising model with a nearest-neighbor interaction:
\begin{equation}\label{eq:XYNN}
    H_{XY,\mathrm{NN}}=\sum_{i=1}^{n}J_{i j}(X_iX_{i+1}+Y_iY_{i+1})+B_Z\sum_{i=1}^{n}Z_i,
\end{equation}
where we set $J_{ij}=\delta_{i,j-1}$ and $B_Z=0.2$; (2) the randomized homogeneous model:
\begin{equation}\label{eq:randomhomo}
    H_{\mathrm{random}}=\sum_{i=1}^{n-1}J_{i j}(X_iX_{i+1}+Y_iY_{i+1}+Z_iZ_{i+1}),
\end{equation}
whose coefficients are chosen randomly, $J_{ij} \in [-1, 1]$ for all $ i < j, i,j = 0,1,\ldots,n-1$; (3) a model with an asymmetry along the $X$-direction:
\begin{equation}\label{eq:XYXNN}
    H_{XY(X),\mathrm{NN}}=\sum_{i=1}^{n}J_{i j}(X_iX_{i+1}+Y_iY_{i+1})+B_X\sum_{i=1}^{n}X_i,
\end{equation}
with $J_{ij}=\delta_{i,j-1}$ and $B_X=0.2$; and (4) the $XYZ$ model with a nearest-neighbor interaction:
\begin{equation}\label{eq:XYZNN}
    H_{XYZ,\mathrm{NN}}=\sum_{i=1}^{n-1}J_{i j}(X_iX_{i+1}+Y_iY_{i+1}+Z_iZ_{i+1}),
\end{equation}
with $J_{ij}=\delta_{i,j-1}$.

For Hamiltonian models $H_{\mathrm{random}}$, $H_{XY(X),\mathrm{NN}}$, and $H_{XYZ,\mathrm{NN}}$, we carry out a similar procedure as for $H_{XY,\mathrm{power}}$ to design an EW with a moderate size. Because of the randomness in the coefficients of $H_{\mathrm{random}}$, we randomly sample its coefficients at each qubit number and evaluate the average performance over the samples. For each Hamiltonian model, we heuristically set the largest EW size as a polynomial in the qubit number [listed in Fig.~\ref{fig_poly_rela_2p}(a)]. The maximal effective entanglement detection times are shown in Fig.~\ref{fig_poly_rela_2p}(b), which remain reasonably steady as the system size increases. We provide details of the data points in SM Sec.~IV.

\begin{figure}[hbtp!]
    \centering
    \includegraphics[width=0.83\columnwidth]{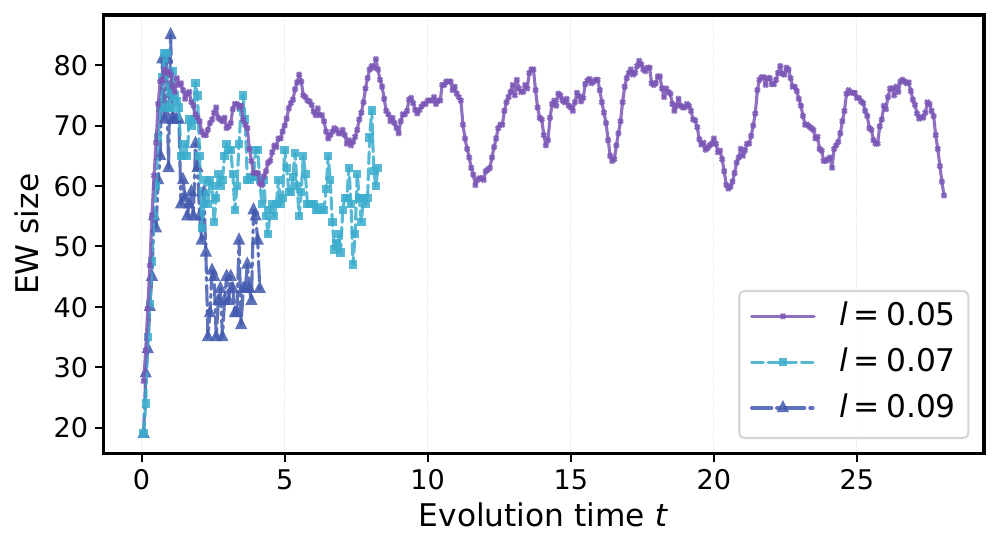}
    \caption{EW-size evolution for $H_{XY,\mathrm{NN}}$. We show a system with 7 qubits. Under the same settings of $t_0$ and $r$, trajectories are plotted for three distinct truncation threshold values. Each simulation stops when the constructed EW can no longer detect entanglement.}
    \label{fig:XYNNEWsize}
\end{figure}

Notably, compared to the other models, the Hamiltonian model $H_{XY,\mathrm{NN}}$ in Eq.~\eqref{eq:XYNN} enjoys a relatively high symmetry. Consequently, we are able to detect entanglement in its evolution for a much longer time with a reasonable EW size. By applying Campbell-Baker-Hausdorff identity, we find that the Pauli-basis decomposition of the EW is supported on a small fixed set of operators; thus with a smaller truncation threshold value, we are guaranteed to achieve a longer entanglement detection time (Fig. \ref{fig:XYNNEWsize}). The coefficients in the Pauli-basis decomposition of the EW can be calculated analytically, and we leave the details in Appendix~B.

In addition to detecting the existence of entanglement, we further design GME witnesses in a similar way. In the simulations, we consider system sizes between $6$ and $11$ qubits and partition the system into three geometrically local parts containing equal or nearly equal numbers of qubits. We demonstrate the performance of the optimized GME witnesses for all Hamiltonian models except $H_{XY,\mathrm{NN}}$ in Fig.~\ref{fig_poly_rela_2p}(c)(d), which are evaluated in a similar fashion as in subfigures (a) and (b). While constraining the GME witness sizes to remain moderate [Fig.~\ref{fig_poly_rela_2p}(c)], as shown in Fig.~\ref{fig_poly_rela_2p}(d), the GME detection times remain reasonably long as the system size grows. For the Hamiltonian model $H_{XY,\mathrm{NN}}$, we are able to carry out an analytical analysis for GME detection analogously to the bipartite-entanglement case. We leave the details of its GME witness design to Appendix~B.

\emph{Outlook. ---}
In this work, we propose an algorithm for designing efficient linear EWs to detect entanglement and GME in the dynamics of general Hamiltonian models. Inspired by generalized stabilizer theory and digitization methods in Hamiltonian simulation, our approach is largely symbolic and remains tractable over long evolution times. In particular, we introduce an effective truncation technique that effectively controls the size of the EWs as the evolution time increases.

Generally, accurately tracking the dynamics of Hamiltonian models becomes increasingly challenging as the system size grows. Our major observation is that for specific tasks such as entanglement detection, it is sufficient to simulate the state only approximately, which can substantially reduce the computational resources.
From the perspective of entanglement detection in large systems, most existing approaches are restricted to fixed symmetric states or special classes such as stabilizer states. Our method provides a new avenue to probe and monitor realistic systems undergoing general Hamiltonian evolutions. In addition, the numerical results for $H_{XY,\mathrm{NN}}$ suggest that entanglement in dynamical systems with highly symmetric Hamiltonians may exhibit distinctive features. We anticipate potential connections between these features and novel physical phenomena specific to such models.
Finally, we expect that our results will stimulate further investigations into the characterization and exploitation of large quantum systems.

\section*{End Matter}
\emph{Appendix A: Algorithm of Designing a Linear EW.---}
We summarize our algorithm to design and heuristically optimize an effective EW. Given an initial state $\ket{\psi_0}$ and the ideal Hamiltonian $H$, we aim to detect entanglement in the realistic system after it undergoes an evolution after time $t_F$. Algorithm~\ref{alg_t_final} is targeted at designing a linear EW for this purpose, of which the size is below a threshold $T$. This algorithm is composed of several smaller modules, which we explain in more detail.

\emph{Module 1:} In this module, we initialize the algorithm parameters in designing and optimizing the linear EW. Specifically, we shall assign the stabilizers of the initial state $\tilde{S}_i^{(0)}$, digitize the Hamiltonian evolution in each unit time $t_0$, and set a set of parameters $l_{\mathrm{small}},l_{\mathrm{large}},l$ for the truncation of less significant Pauli strings in the Pauli-basis decomposition of the evolved stabilizers. Note that the target evolution time $t_F$ might not be an integer multiple of $t_0$, and we hence introduce a residue time term $t_R$. We also initialize two flag variables $x$ and $\mathrm{status}_{\mathrm{last}}$, which will be used for the termination of the algorithm. The flag values upon termination are determined by whether the constructed Hermitian operator can detect the entanglement in the ideal target state $\rho_{\mathrm{ideal}}$ within the size threshold $T$.

\emph{Module 2:} In the second module, we update the stabilizers during Hamiltonian evolution and attempt to construct a linear EW from them. Through lines 9 to 15, we symbolically update the evolved stabilizers approximately. After each unit time $t_0$ and at the end of the evolution $t_F$, we apply a truncation and discard the less significant terms in the Pauli-basis decomposition of the summation of the approximate stabilizers. Suppose the summation of the approximated stabilizers is given by 
\begin{equation}
  \sum_{i=1}^{n}\tilde{S_i}^{(t)} = \sum_{j}c_j\bigotimes_{k=1}^{n}\sigma_j^{(k)}.
\end{equation}
Then, we perform a truncation and discard all the terms in the decomposition with a negligible coefficient:
\begin{equation}\label{eq:trunc}
  \sum_{i=1}^{n}\tilde{S_i}^{(t)}\approx\sum_{\substack{j: \\|c_j|>l\max_s|c_s|}}c_j\bigotimes_{k=1}^{n}\sigma_j^{(k)},
\end{equation}
with $l\in(0,1)$ the truncation threshold value. Afterwards, we solve the SDP of 
\[
    \max_{\sigma\in\mathrm{PPT}}\beta=\tr\left(\sigma\sum_{i=1}^{n}\tilde{S}_i^{(t_F)}\right)
\]
and attempt to construct a linear EW of $W$ (lines 16 to 33). There are several possibilities about this Hermitian operator $W$:
\begin{itemize}
    \item Case 1: The operator $W$ is an EW that can detect entanglement in $\rho_{\mathrm{ideal}}$ with size $x\leq T$ below the pre-determined threshold. In this case, we set $\mathrm{status}_{\mathrm{last}}=0$ and return this valid EW.
    \item Case 2: The operator $W$ is a valid EW, yet its size is larger than $T$. In this case, we shall set $\mathrm{status}_{\mathrm{last}}=1$and adjust $l_0$ to be larger for a more aggresive truncation.
    \item Case 3: The operator $W$ fails to detect entanglement in $\rho_{\mathrm{ideal}}$. In this case, we shall set $\mathrm{status}_{\mathrm{last}}=2$and adjust $l_0$ to be smaller for a more conservative truncation.
\end{itemize}
For cases 2 and 3, the truncation threshold value $l$ is adjusted via adapted $l_{\mathrm{large}}$ and $l_{\mathrm{small}}$, and the algorithm is repeated from the beginning of Module 2 for another attempt to construct an EW. To avoid a non-halting algorithm, we set another termination criterion as $l_{\mathrm{large}}-l_{\mathrm{small}}\leq\epsilon_l$. If such termination condition is met, then the algorithm will output that no EW is found.

\begin{algorithm}[H]

\caption{Design of linear EW}
\label{alg_t_final}
\begin{algorithmic}[1]
    \Require 
    \Statex Hamiltonian $H$
    \Statex initial state $\ket{\psi_0}$ stabilized by $\{S_i\}_{i=1}^{n}$
    \Statex target evolution time $t_F$, unit time $t_{0}$
    \Statex step number $r$
    \Statex initial truncation threshold $l_0$, truncation threshold $l_c$, truncation threshold resolution $\epsilon_l$
    \Statex EW size threshold $T$ 
    \Ensure 
    \Statex EW $W$ w.r.t. $e^{-iHt_{F}}\ket{\psi_0}$ with size less than $T$
    \Statex or $\mathsf{No\,EW\, Found }$
    
    \vspace{2mm}
    \Statex \textbf{Module 1: Parameter initilization}
    \vspace{2mm}
    \State Initialization: $\tilde{S_i}^{(0)} \gets S_i$, $l_{\text{small}} \gets 0$, $l \gets l_0$, $l_{\text{large}} \gets l_c$
    \State $\rho_{\text{ideal}} \gets e^{-iHt_{F}}\ket{\psi_0}$
    \State $k \gets \lfloor t_{F}/t_{0} \rfloor$
    \State $t_{R} \gets t_{F} - k t_0$
    \State $\tilde{U} \gets [I - iHt_{0}/r + (Ht_{0}/r)^2/2]^r$
    \State $\tilde{U}' \gets [I - iHt_{R}/r + (Ht_{R}/r)^2/2]^r$
    \State $x \gets 0$, $\mathrm{status}_{\mathrm{last}} \gets 2$
    \Comment{flag variables}
    
    \vspace{2mm}
    \Statex \textbf{Module 2: Adaptive stabilizer update}
    \vspace{2mm}
    \Repeat
    \For{$\alpha = 1$ to $k$}
        \State $\sum_{i=1}^{n}\tilde{S_i}^{(\alpha t_0)} \gets \tilde{U}\sum_{i=1}^{n}\tilde{S_i}^{((\alpha-1)t_0)} \tilde{U}^\dag$
        \State Truncate $\sum_{i=1}^{n}\tilde{S_i}^{(\alpha t_0)}$ following Eq.~\eqref{eq:trunc}
        \State $\alpha \leftarrow \alpha + 1$
    \EndFor
    \State $\sum_{i=1}^{n}{\tilde{S_i}}^{(t_F)} \gets \tilde{U}' \tilde{S_i}^{(kt_0)}  (\tilde{U}')^{\dagger}$
    \State Truncate $\sum_{i=1}^{n}\tilde{S_i}^{(t_F)}$ following Eq.~\eqref{eq:trunc}
    \State Solve $\beta$ for an EW $W$ w.r.t. $\rho_{\mathrm{ideal}}$
    \If{$\beta$ found}
    \State $W \gets \beta I -\sum_{i=1}^{n}\tilde{S_i}^{(t_F)}$
    \State $x \gets $ size of W
    \If{$\Tr(W\rho_{\text{ideal}})<0$}
    \State $\text{status}_{\text{last}} \gets 0$
    \State $l_{\text{small}} \gets l$
    \Else 
    \State $\text{status}_{\text{last}} \gets 1$
    \State $l_{\text{large}} \gets l$
    \EndIf
    \Else
    \State $\text{status}_{\text{last}} \gets 2$
    \State $l_{\text{large}} \gets l$
    \EndIf
    \State $l \gets (l_{\text{small}}+l_{\text{large}})/2$
    \Until{($\text{status}_{\text{last}} =  0 $ and $x \le T$) or ($\text{status}_{\text{last}} =  1 $ and $x > T$)  or $l_{\text{large}} - l_{\text{small}} \le \epsilon_l$ } 

    \vspace{2mm}
    \Statex \textbf{Module 3: Termination}
    \vspace{2mm}
    \If{$\text{status}_{\text{last}} = 0$ and $x \le T$}
    \Return $W$
    \Else
    \Return ``$\mathsf{No\,EW\, Found }$''
    \EndIf
\end{algorithmic}
\end{algorithm}

We make a remark on how we determine if we find a valid EW. Specifically, we evaluate if our constructed Hermitian operator can detect entanglement in the ideal evolved state, $\rho_{\mathrm{ideal}}$. We calculate $\rho_{\mathrm{ideal}}$ numerically from $e^{-iHt_F}\ket{\psi_0}$ and check if $\tr(W\rho_{\mathrm{ideal}})<0$. The numeric matrix multiplication has a complexity of $O(n^3)$ with $n$ the matrix dimension (advanced numerical algorithms have achieved $O(n^{2.37})$), which is comparable to matrix diagonalization in obtaining the value of $\beta$. While $\rho_{\mathrm{ideal}}$ is numerically calculated, note that it is not immediate to see how one can set up an experimental scheme to detect its entanglement.

\begin{figure}[htbp!]
    \centering
    \includegraphics[width=\textwidth]{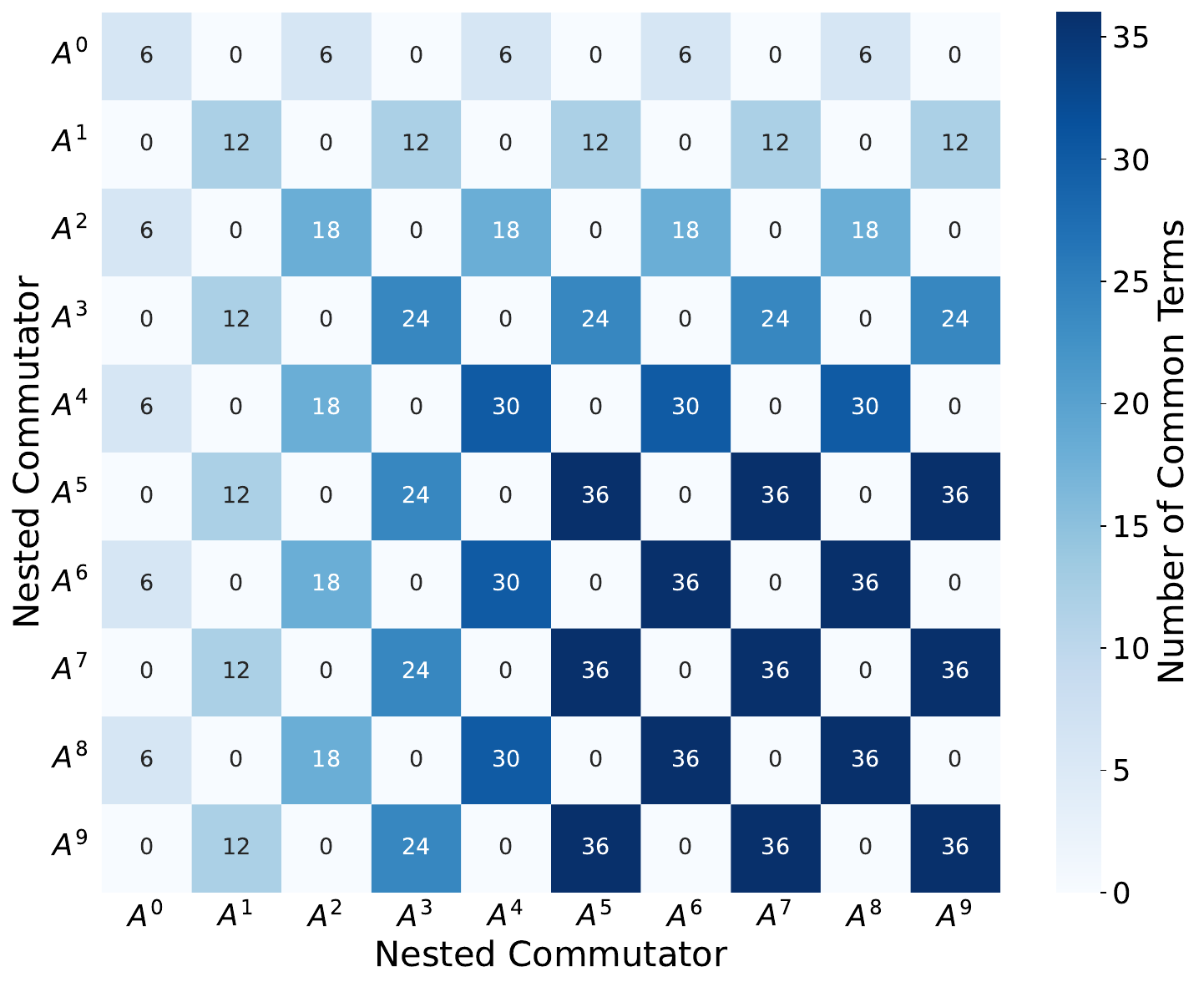} 
    \caption{Overlap matrix of nested commutators $A^{(i)}$ and $A^{(j)}$ for $i,j \in \{0,1,\dots,8\}$ for a system of $6$ qubits, where the overlap is defined as the number of common terms in the Pauli-basis decompositions of $A^{(i)}$ and $A^{(j)}$.}
    \label{Fig:heatmap}
\end{figure}

\emph{Appendix B: Analytical Derivation for Highly Symmetric Hamiltonian Evolution.---}
In some cases with high symmetry, such as the nearest-neighbour transverse-field Ising model $H_{XY,\mathrm{NN}}$ with initial state $\ket{01010\cdots}$, it is possible to calculate the exact evolved stabilizers $\sum S_{i}^{(t)}$ with a low computation cost. Recall the Campbell-Baker-Hausdorff (CBH) identity, which provides the expansion for conjugation of an operator by an exponential:
\begin{equation}
\begin{split}
  e^{-X} Y e^X = Y - [X,Y] + \frac{1}{2!}[X,[X,Y]]  \\- \frac{1}{3!}[X,[X,[X,Y]]] + \cdots.
\end{split}
\end{equation}
By applying the CBH identity to $\sum_i S_i$, we have
\begin{equation}\label{eq:exact_evolsumS}
\begin{split}
\sum_i S_{i_{\text{evol}}} 
= &e^{-iHt} \left(\sum_k S_k\right) e^{iHt} \\
= &\sum_k S_k - it\left[H,\sum_k S_k\right] \\ &+ \frac{(-it)^2}{2!}\left[H,\left[H,\sum_k S_k\right]\right] \\ &+\frac{(-it)^3}{3!}\left[H,\left[H,\left[H,\sum_k S_k\right]\right]\right] + \cdots.
\end{split}
\end{equation}
We can then determine the exact Pauli-basis expansion of the EW by analyzing the nested commutator terms, which we denote as
\begin{align}
A^{(0)} &= \sum_k S_k, \\
A^{(1)} &= \left[H, \sum_k S_k\right], \\
A^{(2)} &= \left[H, \left[H, \sum_k S_k\right]\right], \\ \nonumber
&\vdots \\ 
A^{(k)} &= \underbrace{[H, [H, \ldots [H}_{k \text{ times}}, \sum_k S_k] \ldots ]].
\label{nested_commutators}
\end{align}
For Hamiltonian $H_{XY,\mathrm{power}}$ with initial state $\ket{01010\cdots}$, we find that Eq.~\eqref{eq:exact_evolsumS} is given by a relatively small set of Pauli strings. For example, in a system with six qubits, the EW expansion contains at most 72 Pauli strings at any time. 
More specifically, we list the number of Pauli strings in the expansions of the first nine nested commutators in Fig.~\ref{Fig:heatmap}. We also list the number of Pauli strings that are overlapped between different nested commutators overlap between them. Notably, $A^{(5)}$ and $A^{(7)}$ are expanded into the same set of Pauli strings, and they contain all the Pauli strings in the former odd-indexed $A^{(i)}$. Similarly, $A^{(6)}$ and $A^{(8)}$ are expanded into the same set of Pauli strings, and they contain all the Pauli strings in the former even-indexed $A^{(i)}$. 
By induction, the Pauli strings for $A^{(i)}$ and $A^{(i+2)}$ are the same for $i \geq 5$, and the total number of Pauli strings in the Pauli-basis expansion of $\sum_i S_{i}^{(t)}$ is fixed to be 72.

Since the Pauli-basis expansion of $\sum_i S_{i}^{(t)}$ is within a fixed set of Pauli strings at any time, we can further compute their coefficients through a transformation matrix. For simplicity, denote $S_{\text{sum}}(t)\equiv\sum_i S_{i}^{(t)}$ and write it as
\begin{equation}
    S_{\text{sum}}(t) = \sum_{\alpha=1}^{72} c_\alpha(t) P_\alpha\equiv \overrightarrow{c(t)} \cdot \overrightarrow{P},
\end{equation}
where $\{P_\alpha\}$ is the set of 72 Pauli terms in $S_{\text{sum}}$ and $c_\alpha(t)$ are the corresponding coefficients at time $t$. At $t=0$, $c_\alpha(0)$ are the coefficients of the stabilizers of the initial state. By differentiating $S_{\text{sum}}(t) = e^{-iHt} S_{\text{sum}}(0) e^{iHt}$ with respect to $t$, we have
\begin{equation}
    \frac{d}{d t} S_{\text{sum}}(t)=-i[H, S_{\text{sum}}(t)].
\end{equation}
Because the span is closed under commutation with $H$, the evolution of $\vec{c}$ can be described by a $72 \times 72$ complex matrix $M$ such that the coefficient vector $\overrightarrow{c(t)} \in \mathbb{C}^{72}$ satisfies the linear system $\dot{\overrightarrow{c(t)}}=M \overrightarrow{c(t)}$, where $\overrightarrow{c(0)}$ is known from $S_{\text{sum}}(0)$, and the matrix $M$ can be obtained by calculating the commutators for each Pauli term with the Hamiltonian $H$:
\begin{equation}
    -i\left[H, P_\beta\right]=\sum_\alpha M_{\alpha \beta} P_\alpha,\forall\beta.
\end{equation}
Thus, the solution for the coefficient vector $\overrightarrow{c(t)}$ is given by
\begin{equation}
    \overrightarrow{c(t)} = e^{M t} \overrightarrow{c(0)}.
\end{equation}

\appendix
\section*{Supplementary Material}
\section{Basic notions}
In this section, we review a few concepts of entanglement and stabilizer theory utilized throughout this work. 

For a pure state $\ket{\psi}_{AB}$ defined on a bipartite system $AB$, the entanglement between subsystems $A$ and $B$ is quantified by the entanglement entropy, $E(\ket{\psi})\equiv-\tr(\rho_A\log\rho_A)$, where $\rho_A=\tr_B(\ketbra{\psi})$ is the reduced density matrix of subsystem $A$, and $S(\rho)$ represents the von Neumann entropy of a density operator $\rho$. To extend this quantification to a general mixed state $\rho_{AB}$, we employ the convex-roof construction:
\begin{equation}
    E(\rho)=\min_{\{p_i,\ket{\psi_i}\}}\sum_ip_iE(\ket{\psi_i}),
\end{equation}
where the minimization is taken over all possible convex decompositions satisfying $\rho=\sum_ip_i\ketbra{\psi_i}$. This entanglement measure is conventionally referred to as the entanglement of formation.

Stabilizer theory is traditionally formulated around the structure of the Pauli group. Consider an $n$-qubit Pauli string, $S_i=\bigotimes_{j=1}^{n} P_j$, where each $P_j\in\mathbb{P}$, and $\mathbb{P}$ denotes the single-qubit Pauli group:
\begin{equation}
    \mathbb{P}=\{\pm I,\pm X,\pm Y,\pm Z,\pm iI,\pm iX,\pm iY,\pm iZ\},
\end{equation}
with $X,Y$, and $Z$ representing the non-identity Pauli operators. An eigenvector of $S_i$ associated with an eigenvalue of $+1$ is said to be stabilized by $S_i$. For a set of $n$ mutually commuting and independent Pauli strings, the intersection of their stabilized subspaces uniquely determines a single quantum state $\ket{\psi}$. In this context, two Pauli strings $S_i$ and $S_j$ are independent if they satisfy $\tr(S_iS_j)=0$. Furthermore, the $n$ operators $\{S_1,\cdots,S_n\}$ generate an Abelian group under standard operator multiplication, denoted as $\mathcal{S}_{\psi}=\langle S_1,\cdots,S_n\rangle$. 

As applied in the main text, for a general $n$-qubit state $\ket{\phi}$, one can always find $n$ mutually independent $n$-qubit operators $\{S_i\}_{i=1}^{n}$ that satisfy the stabilizer condition $S_i\ket{\phi}=\ket{\phi}$. However, these generalized operators $\{S_i\}_{i=1}^{n}$ are not restricted to being Pauli strings.

\section{Truncation error analysis}\label{app:analytical}
In this section, we analytically evaluate the approximation errors introduced during the calculation of evolved stabilizers. These errors stem from two primary sources: the truncation of the Taylor series expansion and the discarding of less significant terms in the Pauli-basis decomposition of the stabilizers. Furthermore, we provide a numerical discussion on how these combined truncation errors ultimately affect the final entanglement witness (EW).

\subsection{Truncated Taylor expansion error}
In the Hamiltonian evolution within a unit time, we approximate the evolution operator $e^{-iHt}$ by the truncated Taylor expansion.
The upper bound of the Taylor truncation error and the required truncation series order $K$ were derived in Ref.~\cite{faehrmannShorttimeSimulationQuantum2025}.
By Taylor's remainder theorem, for the the truncated Taylor series of the evolution operator, the operator norm of the remainder can be uppder bounded: 
\begin{equation}
    \norm{\sum_{k=K+1}^{\infty} \frac{(-iHt)^k}{k!}}
    \le \frac{(\norm{H}t)^{K+1}}{(K+1)!}.
\end{equation}
Then, by setting the right hand side to be smaller than an error tolerance $\epsilon$,
we can derive the required series order $K$.
Assuming the scaled evolution time is short such that $\norm{Ht}\le 1$, we have the scaling
\begin{equation}
    K=\mathcal{O}\qty(\frac{\log(1/\epsilon)}{\log\log(1/\epsilon)}),
\end{equation}
which is analogous to the scaling in Ref.~\cite{berrySimulatingHamiltonianDynamics2015}.
When using $r$ segments with $r\ge \norm{H}t$ 
(each segment error is smaller than $\epsilon/r$),
then $K=\mathcal{O}\qty(\frac{\log(r/\epsilon)}{\log\log(r/\epsilon)})$.

Given the tail bound for the truncated evolution operator, 
we next give the error bound of the evolved observable $O(t)$.
Ref.~\cite{faehrmannShorttimeSimulationQuantum2025} leveraged the 1-norm scaling lemma of the Trotter error studied in \cite{childsTheoryTrotterError2021}.
This theorem bounds the tail $\mathbf{R}_K$ of product formulas above the $K$'th order via
\begin{equation}
\begin{split}
    \norm{\mathbf{R}_K \qty(\prod_{l=1}^{L_H} e^{ia_l H_lt})} &= 
    \norm{e^{iHt}-\prod_{l=1}^{L_H} e^{ia_l H_lt}} \\
    &= \frac{(\sum_{l=1}^{L_H}\norm{\alpha_l H_l}t)^{K+1}}{(K+1)!}.
\end{split}
\end{equation}
Consequently, we can bound the error in the truncated evolved observable: 
\begin{align}
    \norm{e^{-iHt} O e^{iHt} - \tilde{O}_K(t)} 
    &\le \norm{\mathbf{R}_K(e^{-iHt}O e^{iHt}) } \\
    &= \norm{\mathbf{R}_K(e^{-iHt}e^{\ln O} e^{iHt}) } \\
    &= \norm{O} \frac{(2\norm{H}t)^{K+1}}{(K+1)!},
\end{align}
where $O$ is lifted to fit the product formula form.

\subsection{Pauli truncation error}
The Pauli truncation error can be bounded as per the framework of Ref.~\cite{begusicRealTimeOperatorEvolution2025}.
We truncate the Pauli operators with small coefficients below a given threshold $\delta$ during evolution.
Let the evolved operator $O_t=\sum_P a_P P$ and the scaled Frobenius norm $\frac{1}{2^n}\Tr(O_0^\dagger O_0)=c_0$.
Then, we have 
\begin{equation}
    \frac{1}{2^n}\Tr(O_t^\dagger O_t) = \sum_P \abs{a_P}^2 \le c_0.
\end{equation}
As $\abs{a_P}\ge \delta$, then
\begin{equation}
    \sum_P \abs{a_P}^2 \ge \sum_P \delta^2 = N\delta^2,
\end{equation}
which yields an upper bound on the number of Pauli operators $\tilde{N}$ for the threshold $\delta$
\begin{equation}
    \tilde{N}\le \frac{c_0}{\delta^2}.
\end{equation}

\subsection{Cancellation of truncation errors in the EW}
Crucially, for the specific purpose of designing entanglement witnesses, we observe that a significant portion of the approximation errors incurred while calculating individual evolved stabilizers effectively cancels out when the operators are summed. 

To illustrate this effect, we consider the power-law Hamiltonian $H_{XY,\mathrm{power}}$ [Eq.~(4) of the main text] in a six-qubit simulation. We compare two distinct error metrics: the cumulative individual error $\sum_i\|S_i-\tilde{S_i}\|$, which characterizes the standard approximation error encountered in conventional state simulations, and the collective error $\|\sum_i(S_i-\tilde{S_i})\|$, which represents the actual physical deviation in the EW constructed from realistic, truncated stabilizers. The resulting values are compiled in Table~\ref{tab:pauli_error}. The comparison demonstrates that the collective approximation error in the constructed EW is significantly lower than the sum of the individual stabilizer errors, confirming that the global observable benefits from error cancellation.

\begin{table}[htbp!]
\caption{Approximation errors in digitized stabilizer evolution with time $t_0$. The first row lists the number of steps that a unit time $t_0$ is divided into.}
\begin{ruledtabular}
\begin{tabular}{c c c}
\textbf{$\#$ Steps} & $\sum_i\|S_i-\tilde{S_i}\|$ & $\|\sum_iS_i-\tilde{S_i}\|$ \\
\hline
1 & $0.70$ & $0.38$ \\
2 & $0.17$ & $0.09$ \\
3 & $0.08$ & $0.04$ \\
5 & $0.03$ & $0.01$ \\
7 & $0.01$ & $7.50\times 10^{-3}$ \\
10 & $6.77\times10^{-3}$ & $3.74\times10^{-3}$ \\
15 & $3.10\times10^{-3}$ & $1.79\times10^{-3}$
\end{tabular}
\end{ruledtabular}
\label{tab:pauli_error}
\end{table}

\section{Effects of choice of truncation threshold}\label{app:TruncTreshold}
In this section, we investigate how the truncation threshold $l$ affects the maximally achievable time for detecting entanglement during Hamiltonian evolution. As a representative example, we consider the power-law XY Hamiltonian model $H_{XY,\mathrm{power}}$ with a system size of 8 qubits. To ensure a consistent comparison, the initial state, the unit time $t_0$, and the number of steps $r$ within the digitized unit-time evolution are kept fixed across all values of $l$. Our analysis focuses on three main attributes: the size of the EW for the evolved state, the distribution of the EW's coefficients in the Pauli basis, and the maximum absolute values among these coefficients. Throughout this discussion, quantities relating to the EW refer directly to the operator $\sum_{i=1}^{n}\tilde{S_i}^{(t)}$, where the identity operator is omitted.

\begin{figure}[H]
\centering
\includegraphics[width=0.85\columnwidth]{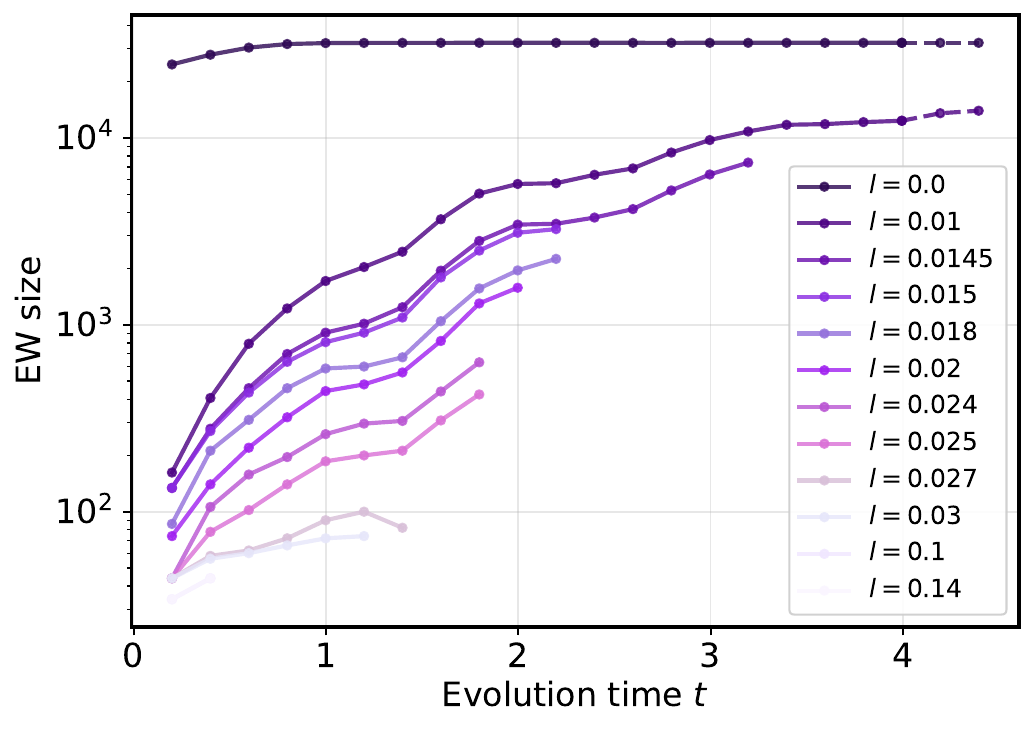} 
\caption{EW-size evolution for $H_{XY,\mathrm{power}}$ in a system of 8 qubits. For each truncation threshold $l$, the EW is constructed at intervals of $4t_0$, and the simulation terminates when the constructed witness can no longer detect entanglement. For $l = 0$ and $l = 0.01$, the curves end with dotted lines to indicate that entanglement detection continues beyond the displayed time frame.}
\label{Fig:EW performance with diff limit}
\end{figure}

In Fig.~\ref{Fig:EW performance with diff limit}, we depict the changes in the EW size during the Hamiltonian evolution for 12 selected values of $l$. For each threshold $l$, the calculation is terminated at the maximum achievable entanglement detection time (in our optimization), defined as the point beyond which the algorithm can no longer output an effective EW to verify the system's entanglement. Notably, for sufficiently small truncation thresholds, entanglement remains detectable over long timescales, even though the EW eventually expands to encompass nearly all terms in the Pauli basis. For instance, when $l =0.01$, our algorithm successfully outputs an effective EW after $1,000$ unit time periods. We indicate this behavior using a dotted line at the end of the displayed time window for $l=0$ and $l =0.01$. As expected, decreasing the truncation threshold extends the maximum entanglement detection time, while the corresponding EW size grows significantly larger.

\begin{figure*}[t]
\centering
\includegraphics[width=0.9\textwidth]{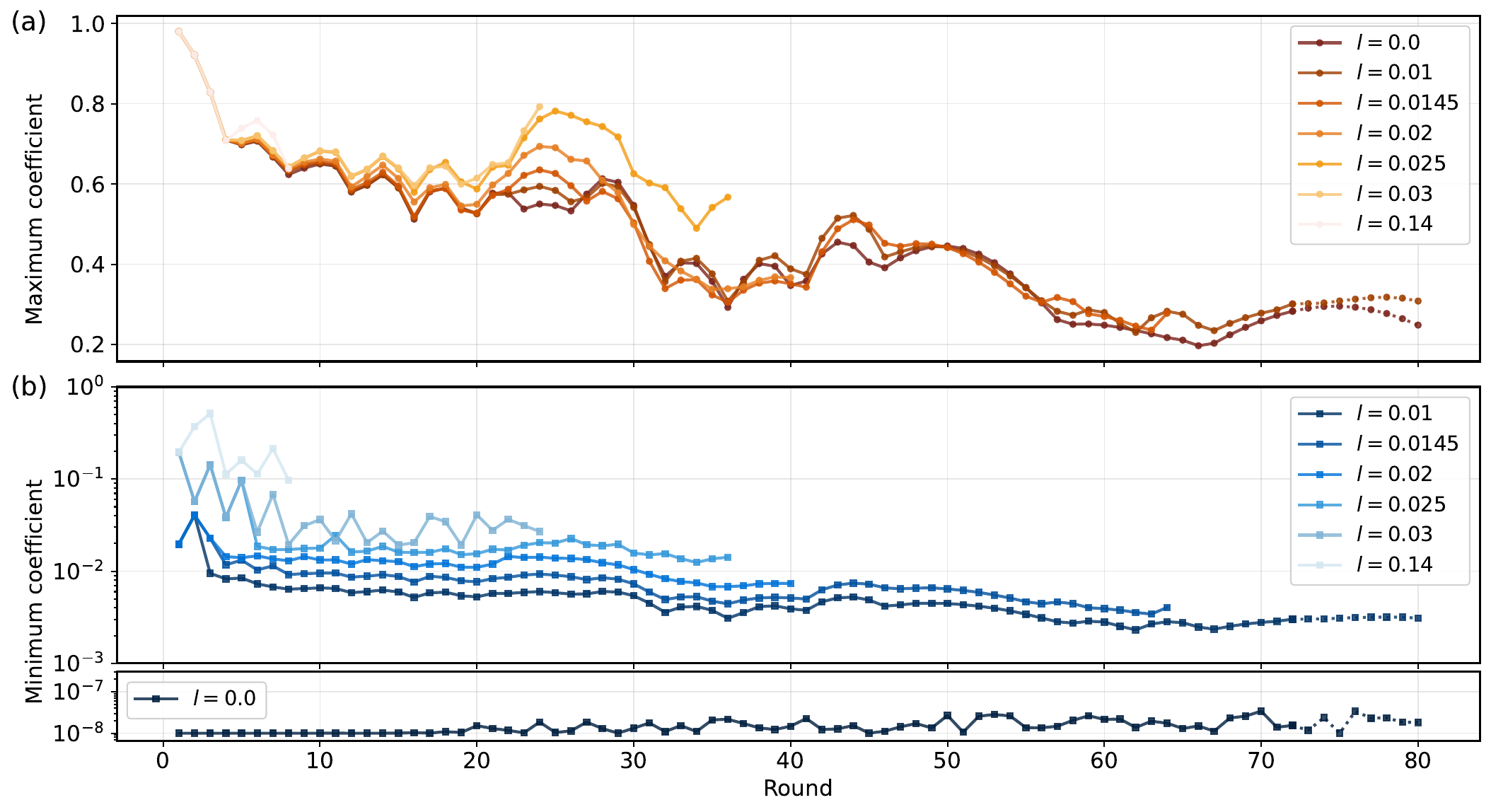}
\caption{Evolution of the maximum and minimum absolute coefficients in the Pauli-basis decomposition of $\sum_{i=1}^{n}\tilde{S_i}^{(t)}$ under $H_{XY,\mathrm{power}}$ for an 8-qubit system. For each truncation threshold $l$, an EW is constructed every $4t_0$, and the operator $\sum_{i=1}^{n}\tilde{S_i}^{(t)}$ is recorded at each unit time step $t_0$. The simulation terminates when the constructed witness can no longer detect entanglement. For $l = 0.0$ and $l = 0.01$, the curves end with dotted lines to indicate that the evolution continues beyond the displayed time frame. Panels depict the (a) maximum absolute coefficient and (b) minimum absolute coefficient among the Pauli terms.}
\label{fig:max min coef evol}
\end{figure*}

Next, we examine the distribution of the coefficients in the Pauli-basis decomposition of the EW. We first depict the changes in the maximum and minimum absolute values of the coefficients in Fig.~\ref{fig:max min coef evol}. At the onset of the Hamiltonian evolution, both values are equal to $1$ for the initial state $\ket{01010\cdots}$, yielding exactly $n$ non-zero terms for an $n$-qubit system. As the system evolves, the maximum coefficient value generally remains higher for larger values of $l$, as shown in Fig.~\ref{fig:max min coef evol}(a); conversely, smaller values of $l$ resolve lower minimum coefficients, as shown in Fig.~\ref{fig:max min coef evol}(b).

\begin{figure*}[t]
\centering
\includegraphics[width=0.95\textwidth]{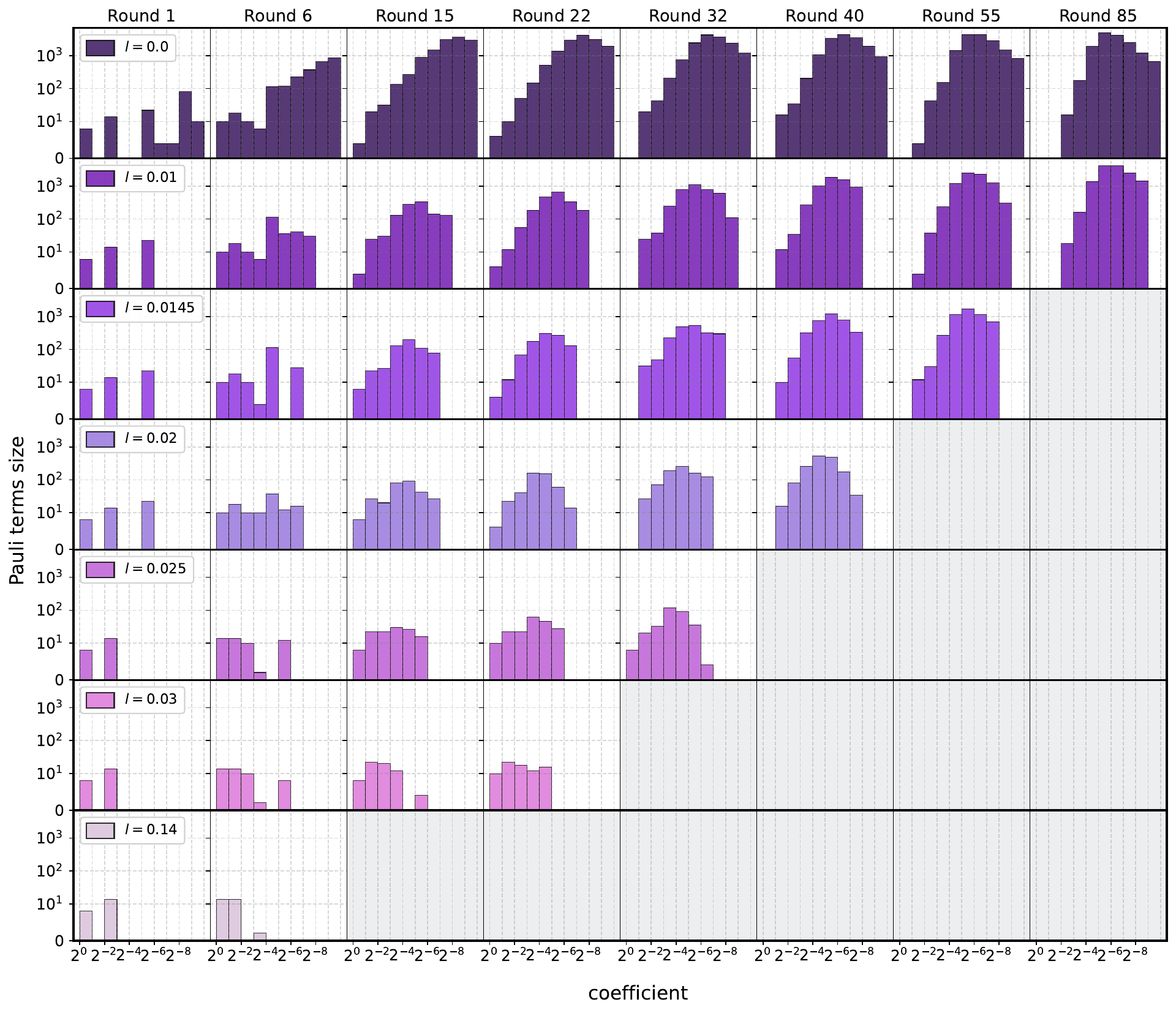}
\caption{Evolution of the Pauli term distribution for $\sum_{i=1}^{n}\tilde{S_i}^{(t)}$ under $H_{XY,\mathrm{power}}$ in an 8-qubit system. The data is displayed across seven truncation thresholds $l$ and eight evolution rounds. For a given $l$ and round, bars represent the number of Pauli terms with coefficients falling within the interval $(2^{-a-1},2^{-a}]$, where the integer $a \in [0,9]$. For convenience, the horizontal axis is reversed. Gray blocks indicate rounds where the constructed EW fails to detect entanglement. The vertical axis utilizes a symmetric logarithmic scale with a linear threshold of 10.}

\label{fig:coeff distr}
\end{figure*}

To complement these observations, Fig.~\ref{fig:coeff distr} tracks the evolution of the coefficient distribution under the same physical settings. Here, as the Hamiltonian model evolves, we plot the number of terms in the EW's Pauli-basis decomposition whose coefficient magnitudes fall into bins defined by $(2^{-a-1},2^{-a}]$, where $a$ is an integer between $0$ and $9$. Each column of the figure represents a given evolution time as the number of rounds of $t_0$. In the baseline case of $l = 0$ where no truncation is performed, more than $15,000$ Pauli strings possess coefficients lying within the lowest resolved region, $(0,2^{-10}]$. In contrast, for all non-zero values of $l$ considered here, no Pauli strings survive in this specific bin. This demonstrates that a smaller truncation threshold allows the EW to retain a vast number of Pauli strings with small absolute coefficients. These terms preserve fine-grained information about the target state during the Hamiltonian evolution, mirroring the physical intuition that an initially peaked coefficient distribution progressively flattens as a complex quantum system evolves.

Finally, we observe that EWs constructed with smaller values of $l$ generally feature fewer Pauli-basis terms with large coefficients. By examining the column-wise coefficient distributions across different truncation thresholds in Fig.~\ref{fig:coeff distr}, it is evident that the height of the first bar, which represents the number of coefficients whose absolute values reside in $(2^{-1},1]$, is lower for smaller values of $l$. This behavior becomes more pronounced from Round 15 onwards, which likely stems from destructive interference mediated by the small-coefficient Pauli terms that are preserved under more stringent truncation thresholds. As evidence of this mechanism, in the first round, the EWs for $l \le 0.02$ contain Pauli-basis terms with coefficients smaller than or equal to $2^{-5}$, whereas all coefficients in the Pauli-basis decomposition exceed this value for $l > 0.02$. A similar suppression is observable at longer evolution times. The absence of such fine-grained information under larger thresholds thus results in the concentration of the coefficient distribution into fewer dominating terms.

\section{Simulation data details}
\label{app:simulation data details}
For completeness, we present the detailed simulation data corresponding to Fig.~3 in the main text from Table~\ref{tab:pow_2p} through Table~\ref{tab:H3_3p}. Specifically, for the implementation of the algorithms optimizing both standard entanglement and genuine multipartite entanglement (GME) detection under the evolution of the Hamiltonian models $H_{XY,\mathrm{power}}$, $H_{XY(X),\mathrm{NN}}$, and $H_{XYZ,\mathrm{NN}}$, we list the key operational parameters for each system size $n$. These parameters include the prescribed truncation thresholds $l$, the number of unit time steps (rounds) over which entanglement remains detectable alongside the corresponding total evolution time $t$, the total number of non-zero terms surviving in the Pauli-basis decomposition after truncation, and the optimal coefficient $\beta$ of the linear EW evaluated at the final instance of valid detection.  For the random Hamiltonian $H_{\mathrm{random}}$, which requires stochastic sampling of the interaction coefficients, we explicitly list the corresponding data for each independent realization, as specified by its unique sample label number.

\begin{table}[H]
\caption{Details of the data points displayed in Fig.~3(a) in the main text for $H_{XY,\mathrm{power}}$. The columns are the number of qubits $n$, truncation threshold $l$, number of $t_0$ evolved (rounds), maximal evolution time $t$, the EW size (the number of Pauli terms), and the value of $\beta$ in the EW at the final time we are able to detect entanglement in running Algorithm~1 in the main text, respectively. Except for $H_{\mathrm{random}}$, other tables are presented in a similar fashion.}
\begin{ruledtabular}
\begin{tabular}{cccccc}
    $n$ & $l$ & rounds & $t$ & Pauli terms & $\beta$ \\
    \hline
    6 & 0.05 & 14 & 1.51 & 184 & 3.243 \\ 
    7 & 0.035 & 23 & 1.67 & 282 & 4.031 \\ 
    8 & 0.025 & 36 & 1.8 & 424 & 3.921 \\ 
    9 & 0.017 & 50 & 1.76 & 736 & 4.801 \\ 
    10 & 0.01 & 40 & 1.01 & 828 & 6.697 \\ 
    11 & 0.008 & 72 & 1.34 & 1242 & 6.636 \\
\end{tabular}
\end{ruledtabular}
\label{tab:pow_2p}
\end{table}

\begin{table}[H]
\caption{Details of the data points displayed in Fig.~3(c) in the main text for $H_{XY,\mathrm{power}}$.}
\begin{ruledtabular}
\begin{tabular}{cccccc}
    $n$ & $l$ & rounds & $t$ & Pauli terms & $\beta$ \\ \hline
    6 & 0.053 & 13 & 1.4 & 170 & 3.082 \\ 
    7 & 0.035 & 20 & 1.45 & 218 & 3.449 \\ 
    8 & 0.025 & 33 & 1.65 & 340 & 3.998 \\ 
    9 & 0.018 & 50 & 1.76 & 426 & 3.832 \\ 
    10 & 0.0129 & 70 & 1.77 & 572 & 4.403 \\ 
    11 & 0.005 & 30 & 0.56 & 916 & 8.34 \\
\end{tabular}
\end{ruledtabular}
\label{tab:pow_3p}
\end{table}

\begin{table}[tbp]
\caption{Data details of sampled Hamiltonians $H_{\mathrm{random}}$ for Fig.~3(a) in the main text. For system sizes of $6,7$, and $10$ qubits, we each sample five sets of coefficients and present the relevant quantities for each sample. Because of the limitation in computational resources, we only sample three Hamiltonians for $n=8,9$ and one Hamiltonian for $n=11$ qubits.} 

\begin{ruledtabular}
\begin{tabular}{ccccccc}
    $n$ & $l$ & sample label & rounds & $t$ & Pauli terms & $\beta$ \\ \hline
    6 & 0.1 & 1 & 21 & 2.27 & 674 & 2.18 \\
    6 & 0.1 & 2 & 9 & 0.97 & 366 & 2.34 \\
    6 & 0.1 & 3 & 6 & 0.65 & 171 & 3.36 \\
    6 & 0.1 & 4 & 3 & 0.32 & 46 & 4.38 \\
    6 & 0.1 & 5 & 0 & 0 & 38 & 5.13 \\
    7 & 0.05 & 1 & 6 & 0.43 & 953 & 4.3 \\
    7 & 0.05 & 2 & 6 & 0.43 & 649 & 3.9 \\
    7 & 0.05 & 3 & 6 & 0.43 & 158 & 4.22 \\
    7 & 0.05 & 4 & 9 & 0.65 & 307 & 3.93 \\
    7 & 0.05 & 5 & 12 & 0.87 & 1622 & 3.11 \\
    8 & 0.04 & 1 & 9 & 0.45 & 1251 & 4.7 \\
    8 & 0.04 & 2 & 15 & 0.75 & 3031 & 3.39 \\
    8 & 0.04 & 3 & 9 & 0.45 & 342 & 5.05 \\
    9 & 0.02 & 1 & 20 & 0.7 & 3918 & 4.36 \\
    9 & 0.02 & 2 & 12 & 0.42 & 2946 & 4.66 \\
    9 & 0.02 & 3 & 12 & 0.42 & 2103 & 4.81 \\
    10 & 0.013 & 1 & 20 & 0.51 & 9000 & 4.93 \\
    10 & 0.013 & 2 & 16 & 0.41 & 4303 & 5.06 \\
    10 & 0.013 & 3 & 12 & 0.3 & 1646 & 6.4 \\
    10 & 0.013 & 4 & 16 & 0.4 & 5325 & 5.98 \\
    10 & 0.013 & 5 & 16 & 0.41 & 8917 & 5.23 \\
    11 & 0.005 & 1 & 18 & 0.33 & 10906 & 7.44 \\
\end{tabular}
\end{ruledtabular}
\label{tab:rand_2p}
\end{table}

\begin{table}[H]
\caption{Details of the data points displayed in Fig.~3(a) in the main text for $H_{\mathrm{random}}$. The third column lists the number of samples evaluated for each given system size and truncation threshold. The remaining columns share the same physical meanings as those in Table~\ref{tab:pow_2p}, but represent values averaged over the random samples.}
\begin{ruledtabular}
\begin{tabular}{ccccccc}
    $n$ & $l$ & counts & rounds & $t$ & Pauli terms & $\beta$ \\ \hline
    6 & 0.1 & 5 & 7.8 & 0.84 & 259 & 3.48 \\
    7 & 0.05 & 5 & 7.8 & 0.56 & 738 & 3.89 \\
    8 & 0.04 & 3 & 11 & 0.55 & 1541 & 4.38 \\
    9 & 0.02 & 3 & 14.67 & 0.51 & 2989 & 4.61 \\
    10 & 0.013 & 5 & 16 & 0.41 & 5838 & 5.52 \\
    11 & 0.005 & 1 & 18 & 0.33 & 10906 & 7.44
\end{tabular}
\end{ruledtabular}
\label{tab:rand_2p_avg}
\end{table}

\begin{table}[hbt!]
\caption{Data details of sampled Hamiltonians $H_{\mathrm{random}}$ for Fig.~3(c) in the main text. The columns represent the same quantities as those in Table~\ref{tab:rand_2p}.}
\begin{ruledtabular}
\begin{tabular}{ccccccc}
    $n$ & $l$ & sample label & rounds & $t$ & Pauli terms & $\beta$ \\ \hline
    6 & 0.12 & 1 & 4 & 0.43 & 94 & 3.37 \\
    6 & 0.12 & 2 & 4 & 0.43 & 49 & 3.39 \\
    6 & 0.12 & 3 & 4 & 0.43 & 79 & 3.34 \\
    6 & 0.12 & 4 & 8 & 0.86 & 327 & 2.52 \\
    6 & 0.12 & 5 & 12 & 1.3 & 440 & 2.19 \\
    7 & 0.058 & 1 & 4 & 0.29 & 83 & 5.2 \\
    7 & 0.058 & 2 & 8 & 0.58 & 614 & 3.11 \\
    7 & 0.058 & 3 & 8 & 0.58 & 458 & 3.68 \\
    7 & 0.058 & 4 & 4 & 0.29 & 152 & 4.19 \\
    7 & 0.058 & 5 & 8 & 0.58 & 379 & 3.86 \\
    8 & 0.04 & 1 & 9 & 0.45 & 709 & 4.11 \\
    8 & 0.04 & 2 & 9 & 0.45 & 1371 & 3.9 \\
    8 & 0.04 & 3 & 9 & 0.45 & 249 & 4.41 \\
    8 & 0.04 & 4 & 6 & 0.3 & 258 & 5.5 \\
    8 & 0.04 & 5 & 9 & 0.45 & 1035 & 4.04 \\
    9 & 0.027 & 1 & 8 & 0.28 & 291 & 6.04 \\
    9 & 0.027 & 2 & 16 & 0.56 & 2961 & 4.02 \\
    9 & 0.027 & 3 & 8 & 0.28 & 385 & 6.02 \\
    9 & 0.027 & 4 & 12 & 0.42 & 1599 & 4.33 \\
    9 & 0.027 & 5 & 12 & 0.42 & 1333 & 4.67 \\
    10 & 0.025 & 1 & 16 & 0.41 & 1887 & 5.16 \\
    10 & 0.025 & 2 & 12 & 0.3 & 192 & 6.6 \\
    10 & 0.025 & 3 & 16 & 0.41 & 4639 & 4.39 \\
    11 & 0.009 & 1 & 15 & 0.28 & 2718 & 7.5 \\
\end{tabular}
\end{ruledtabular}
\label{tab:rand_3p}
\end{table}

\begin{table}[H]
\caption{Details of the data points displayed in Fig.~3(c) in the main text for $H_{\mathrm{random}}$. The data are obtained in a similar manner as in Table~\ref{tab:rand_2p_avg}.}
\begin{ruledtabular}
\begin{tabular}{ccccccc}
    $n$ & $l$ & counts & rounds & $t$ & Pauli terms & $\beta$ \\ \hline
    6 & 0.12 & 5 & 6.4 & 0.69 & 198 & 2.96 \\
    7 & 0.058 & 5 & 6.4 & 0.46 & 337 & 4.01 \\
    8 & 0.04 & 5 & 8.4 & 0.42 & 724 & 4.39 \\
    9 & 0.027 & 5 & 11.2 & 0.39 & 1314 & 5.02 \\
    10 & 0.025 & 3 & 14.67 & 0.37 & 2239 & 5.39 \\
    11 & 0.009 & 1 & 15 & 0.28 & 2718 & 7.5
\end{tabular}
\end{ruledtabular}   
\label{tab:rand_3p_avg}
\end{table}

\begin{table}[H]
\caption{Details of the data points displayed in Fig.~3(a) in the main text for $H_{XY(X),\mathrm{NN}}$.}
\begin{ruledtabular}   
\begin{tabular}{cccccc}
    $n$ & $l$ & rounds & $t$ & Pauli terms & $\beta$ \\ \hline
    6 & 0.2 & 10 & 1.08 & 54 & 3.77 \\
    7 & 0.04 & 36 & 2.61 & 73 & 3.4 \\
    8 & 0.03 & 21 & 1.05 & 112 & 6.58 \\
    9 & 0.015 & 39 & 1.37 & 139 & 6.89 \\
    10 & 0.025 & 57 & 1.44 & 180 & 7.03 \\
    11 & 0.02 & 75 & 1.39 & 213 & 7.66 \\
    12 & 0.015 & 110 & 1.52 & 288 & 8.27
\end{tabular}
\end{ruledtabular}
\label{tab:H1_2p}
\end{table}

\begin{table}[H]
\caption{Details of the data points displayed in Fig.~3(c) in the main text for $H_{XY(X),\mathrm{NN}}$.}
\begin{ruledtabular}
\begin{tabular}{cccccc}
    $n$ & $l$ & rounds & $t$ & Pauli terms & $\beta$ \\ \hline
    6 & 0.2 & 12 & 1.3 & 60 & 3.25 \\
    7 & 0.04 & 38 & 2.75 & 73 & 3.43 \\
    8 & 0.03 & 16 & 0.8 & 112 & 3.77 \\
    9 & 0.02 & 48 & 1.69 & 145 & 4.1 \\
    10 & 0.025 & 60 & 1.52 & 180 & 6.48 \\
    11 & 0.02 & 81 & 1.51 & 217 & 5.58
\end{tabular}
\end{ruledtabular}
\label{tab:H1_3p}
\end{table}

\begin{table}[H]
\caption{Details of the data points displayed in Fig.~3(a) in the main text for $H_{XYZ,\mathrm{NN}}$.}
\begin{ruledtabular}
\begin{tabular}{cccccc}
    $n$ & $l$ & rounds & $t$ & Pauli terms & $\beta$ \\ \hline
    6 & 0.15 & 8 & 0.86 & 334 & 2.43 \\
    7 & 0.08 & 9 & 0.65 & 484 & 3.7 \\
    8 & 0.07 & 12 & 0.6 & 600 & 4.44 \\
    9 & 0.04 & 18 & 0.63 & 757 & 5.4 \\
    10 & 0.035 & 21 & 0.53 & 910 & 6.47 \\
    11 & 0.015 & 27 & 0.5 & 1011 & 8.05 \\
    12 & 0.015 & 36 & 0.5 & 1284 & 8.78
\end{tabular}
\end{ruledtabular}
\label{tab:H3_2p}
\end{table}

\begin{table}[H]
\caption{Details of the data points displayed in Fig.~3(c) in the main text for $H_{XYZ,\mathrm{NN}}$.}
\begin{ruledtabular}
\begin{tabular}{cccccc}
    $n$ & $l$ & rounds & $t$ & Pauli terms & $\beta$ \\ \hline
    6 & 0.15 & 8 & 0.86 & 334 & 2.24 \\
    7 & 0.085 & 9 & 0.65 & 411 & 3.07 \\
    8 & 0.07 & 12 & 0.6 & 600 & 4.22 \\
    9 & 0.04 & 18 & 0.63 & 757 & 4.83 \\
    10 & 0.03 & 21 & 0.53 & 1230 & 6.04 \\
    11 & 0.015 & 30 & 0.56 & 1473 & 6.51
\end{tabular}
\end{ruledtabular}
\label{tab:H3_3p}
\end{table}

\bibliographystyle{apsrev4-2}

\bibliographystyle{truncate_ref}
\bibliography{TrotterEW,ref_aps}

\end{document}